\documentclass[aip,jap,reprint,superscriptaddress]{revtex4-1} 
\usepackage{amsfonts, hyperref}
\usepackage{graphicx}
\begin{document}

\title{Power optimization for domain wall motion in ferromagnetic
  nanowires}

\author{O.~A.~Tretiakov}
\author{Y.~Liu}
\author{Ar.~Abanov}

\affiliation{
            Department of Physics \& Astronomy,
	    Texas A\&M University,
            College Station, Texas 77843-4242, USA
}

\date{October 7, 2010}

\begin{abstract}
The current mediated domain-wall dynamics in a thin ferromagnetic wire
is investigated. We derive the effective equations of motion of the
domain wall. They are used to study the possibility to optimize the
power supplied by electric current for the motion of domain walls in a
nanowire. We show that a certain resonant time-dependent current
moving a domain wall can significantly reduce the Joule heating in the
wire, and thus it can lead to a novel proposal for the most energy
efficient memory devices. We discuss how Gilbert damping,
non-adiabatic spin transfer torque, and the presence of
Dzyaloshinskii-Moriya interaction can effect this power optimization.
\end{abstract}

\maketitle

\textit{Introduction.}  Due to its direct relevance to future memory
and logic devices, the dynamics of domain walls (DW) in magnetic
nanowires has become recently a very popular topic.\cite{Parkin08,
  *Hayashi08, Allwood01, Allwood02} There are mainly two goals which
scientists try to achieve in this field. One goal is to move the
domain walls with higher velocity in order to make faster memory or
computer logic. The other one is inspired by the modern trend of
energy conservation and concerns a power optimization of the
domain-wall devices.

Generally, the domain walls can be manipulated whether by a magnetic
field \cite{Ono99, Allwood02} or electric current. \cite{Yamaguchi04,
  Parkin08, *Hayashi08} Although the latter method is preferred for
industrial applications due to the difficulty with the application of
magnetic fields locally to small wires. For this reason, we consider
in this paper the current induced domain-wall dynamics. We make a
proposal on how to optimize the power for the DW motion by means of
reducing the losses on Joule heating in ferromagnetic
nanowires. \cite{Tretiakov:losses} Moreover, because the averaged over
time (often called drift) velocity of a DW generally increases with
applied current, we also address the first goal. Namely, our proposal
allows to move the DWs with higher current densities without burning
the wire by the excessive heat and thus archive higher drift
velocities of DWs. The central idea of this proposal is to employ
resonant time-dependent current to move DWs, where the period of the
current pulses is related to the periodic motion of DW internal
degrees of freedom.

\begin{figure}
\includegraphics[width=0.9\columnwidth]{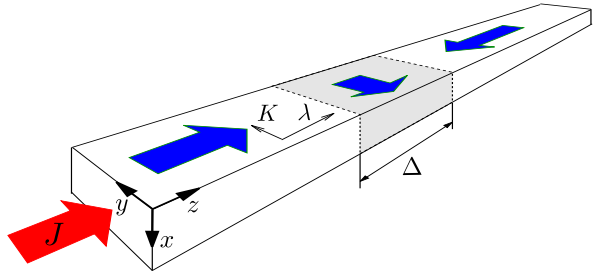} 
\caption{(color online) A schematic view of a current-driven domain
  wall in a ferromagnetic wire. The DW width is $\Delta$.}
\label{fig:DW} 
\end{figure}

The schematic view of a domain wall in a narrow ferromagnetic wire is
shown in Fig.~\ref{fig:DW}. These DWs are characterized by their width
$\Delta$ which is mainly determined by exchange interaction and
anisotropy along the wire $\lambda$. Another important quantity is the
transverse anisotropy across the wire $K$, which governs the pinning
of the transverse component of the DW magnetization. When no current
is applied to the wire it leads to two degenerate positions of the
transverse magnetization component of the wall: as shown in
Fig.~\ref{fig:DW} and anti-parallel to it. 

To describe the dynamics of DW in a thin wire we derived the effective
equations of motion from generalized
Landau-Lifshitz-Gilbert\cite{Zhang04, Thiaville05} (LLG) equation with
the current $J$,
\begin{equation}
  \dot{\mathbf{S}}=\mathbf{S}\times\mathbf{H}_{eff}
-J\frac{\partial\mathbf{S}}{\partial z}
+\beta J\mathbf{S}\times\frac{\partial\mathbf{S}}{\partial z}
+\alpha\mathbf{S}\times\dot{\mathbf{S}},
\label{eq:LLG}
\end{equation}
where $\mathbf{S}$ is magnetization unit vector,
$\mathbf{H}_{eff}=\delta\mathcal{H}/\delta\mathbf{S}$ is the effective
magnetic field given by the Hamiltonian $\mathcal{H}$ of the system,
$\beta$ is non-adiabatic spin torque constant, and $\alpha$ is Gilbert
damping constant.  The derivation of the effective equations of motion
is based on the fact that in thin ferromagnetic wires the static DWs
are rigid topologically constrained spin-textures. Therefore, for not
too strong drive, their dynamics can be described in terms of only a
few collective coordinates associated with the DW degrees of
freedom. \cite{Tretiakov08, *Clarke08} In very thin wires, there are
two collective coordinates corresponding to two softest modes of the
DW motion: the DW position along the wire $z_0$ and the magnetization
angle $\phi$ in the DW around the wire axis. All other degrees of
freedom are gapped by strong anisotropic energy along the wire.

By applying the orthogonality condition to LLG, one can obtain the
equations of motion for the two DW softest modes, $z_0(t)$ and $\phi
(t)$,\cite{Tretiakov_DMI}
\begin{eqnarray} 
\dot{z}_{0}&=& AJ+B[J-j_{c}\sin(2\phi)],
\label{eq:z0Dot}\\
\dot{\phi}&=& C[J-j_{c}\sin(2\phi)]\label{eq:phiDot},
\end{eqnarray}
where $J(t)$ is a time-dependent current. The coefficients $A$, $B$,
$C$, and critical current $j_c$ can be evaluated for a particular
model in terms of $\alpha$, $\beta$ and other microscopic
parameters. Following Ref.~\onlinecite{Tretiakov_DMI}, for the model
with Dzyaloshinskii-Moriya interaction (DMI) one can find $A=
\beta/\alpha$, $B=
(\alpha-\beta)(1+\alpha\Gamma\Delta)/[\alpha(1+\alpha^{2})]$, $C=
(\alpha-\beta) \Delta/[(1+\alpha^{2})\Delta_{0}^{2}]$, and
$j_c=(\alpha K\Delta /\left|\alpha-\beta\right|) [\pi \Gamma\Delta
  /\sinh(\pi\Gamma\Delta)]$, where $J_{\rm{ex}}$ is exchange constant,
$D$ is DMI constant, and $\Gamma =D/J_{\rm{ex}}$. Also,
$\Delta=\Delta_0 /\sqrt{1-\Gamma^2 \Delta_0^2}$ where $\Delta_0$ is
the DW width in the absence of DMI.

Alternatively, Eqs.~(\ref{eq:z0Dot}) and (\ref{eq:phiDot}) can be
obtained in a more general framework by means of symmetry arguments.
We note that because of the translational invariance $\dot{z}_{0}$ and
$\dot{\phi}$ cannot depend on $z_0$. Furthermore, to the first order in
small transverse anisotropy $K$, $\dot{\phi}$ and $\dot{z}_{0}$ are
proportional to the first harmonic $\sin(2\phi)$. Then the expansion
in small current $J$ up to a linear in $J$ order gives
Eqs.~(\ref{eq:z0Dot}) and (\ref{eq:phiDot}). In this case the
coefficients $A$, $B$, $C$, and $j_c$ have to be determined directly
from experimental measurements. \cite{BeachPRL09, LiuEMF}

\begin{figure}
\includegraphics[width=1\columnwidth]{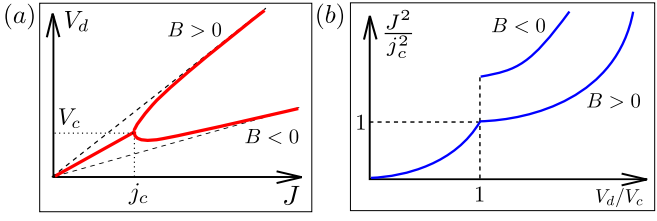}
\caption{(color online) DW motion characteristics for dc currents. (a)
  Drift velocity $V_d$ of DW as a function of current $J$ for $B>0$
  and $B<0$, see Eq.~(\ref{eq:z0Dot}). The slope at $J<j_c$ is given
  by $A$, whereas at $J\gg j_c$ it is $A+B$. (b) Power of Ohmic losses
  $p_{\rm{dc}}(V_d/V_c)=J^2/j_c^2$ as a function of drift velocity
  $V_d$. For $B<0$ the power has a discontinuity at $V_d/V_c =1$.}
\label{fig:domain_wall} 
\end{figure}

 For the dc current applied to the wire the DW dynamics governed by
 Eqs.~(\ref{eq:z0Dot}) and (\ref{eq:phiDot}) can be obtained
 explicitly.\cite{Tretiakov_DMI} For $J<j_c$ and $A\neq 0$ the DW only
 moves along the wire and is tilted on angle $\phi_0$ from the
 transverse-anisotropy easy axis given by condition
 $\sin(2\phi_0)=J/j_c$. The drift velocity is $V_d =\langle
 \dot{z}_{0}(J) \rangle=AJ$, see Eq.~(\ref{eq:z0Dot}). Therefore, the
 linear slope of $V_d(J)$ below $j_c$ gives constant $A$, see
 Fig.~\ref{fig:domain_wall} (a). The value of $j_c$ is determined as
 the endpoint of this linear regime. At $J=j_{c}$ the magnetization
 angle becomes perpendicular to the easy axis, $\phi_0=\pi/2$.  For
 $J>j_{c}$ the DW both moves and rotates, and Eqs.~(\ref{eq:z0Dot})
 and (\ref{eq:phiDot}) give $V_{d} = AJ +B\sqrt{J^{2}-j_{c}^{2}}$, so
 that the slope of $V_d(J)$ at large $J$ gives $A+B$.

\textit{Power optimization.}  The largest losses in the nanowire with
a DW are the Ohmic losses of the current. In general, the influence of
the DW on the resistance is negligible and therefore we can assume
that the resistance of the wire is constant with time.  Then the
time-averaged power of Ohmic losses is proportional to $\langle J^2(t)
\rangle$. Since the resistance is almost constant, in this paper we
will calculate $P=\langle J^2(t) \rangle$ and loosely call it the
power of Ohmic losses. Our goal is to minimize the Ohmic losses while
keeping the DW moving with a given constant drift velocity.

For the following it will be convenient to introduce the dimensionless
variables for time, drift velocity, current, power, and the ratio of
slopes of $V_d(J)$ at large and small currents,
\begin{equation}
\tau=Cj_{c}t,\quad
v_{d}=\frac{V_{d}}{V_{c}},\quad j=\frac{J}{j_c},\quad
p=\frac{\mathcal{P}}{j_{c}^{2}},\quad a=\frac{A+B}{A}.
\label{eq:dimensionless}
\end{equation} 
Although we note that in the special case of $\alpha =\beta$, it can
be shown that $C=B=0$ and one cannot use dimensionless
variables~(\ref{eq:dimensionless}). However, in this case the DW
dynamics is trivial: \cite{Barnes05} the DW does not rotate
$\phi=0,\pi$ and moves with the velocity $\dot{z}_0=J$.

First, we consider the case of dc current and the power as a function
of drift velocity. For $v_{d}<1$ we find $p_{\rm{dc}}=v_{d}^{2}$.  For
currents above $j_c$ the power $p_{\rm{dc}}(v_{d})=j^2$ is given in
terms of drift velocity $v_{d}=j+(B/A)\sqrt{j^{2}-1}$ as shown in
Fig.~\ref{fig:domain_wall} (b). The power is quadratic in $v_d$, and
for $B<0$ it has a discontinuity at $v_d =1$.

In general, the DW motion has some period $T$ and current $j(\tau)$
must be a periodic function with the same $T$ to minimize the Ohmic
losses.  Measuring the angle from the hard axis instead of easy axis
and scaling it by 2, i.e, $2\phi= \theta -\pi/2$, we can write the
dimensionless current drift velocity as \cite{Tretiakov:losses}
\begin{equation}
j(\tau)= \dot{\theta}/2 -\cos\theta,\quad
v_{d}=\frac{a}{2}\langle\dot{\theta}\rangle
-\langle\cos\theta\rangle ,
\label{eq:VdAve}
\end{equation}
where $\dot{\theta}=\partial\theta/\partial\tau$.  

To minimize the power of Ohmic losses we need to find the minimum of
$\langle j^{2}(\tau) \rangle$ at fixed $v_{d}$,
\begin{equation}
\overline{p} = \left\langle
(\dot{\theta}/2-\cos\theta )^{2}
-2\rho (a\dot{\theta}/2
-\cos\theta -v_{d})\right\rangle, 
\label{eq:Power}
\end{equation}
where we use a Lagrange multiplier $2\rho$ to account for the
constraint given by $v_{d}$ from Eq.~(\ref{eq:VdAve}).
Power~(\ref{eq:Power}) can be considered as an effective action for a
particle in a periodic potential $U$, and its minimization gives the
equation of motion $\ddot{\theta}/2=-\partial U/\partial \theta$ which
in turn can be reduced to
\begin{equation}
\dot{\theta} = \pm 2\sqrt{d-U(\theta,\rho)},
\quad U(\theta,\rho)=-\cos^{2}\theta-2\rho\cos\theta.
\label{eq:thetaDotD}
\end{equation}
where $d$ is an arbitrary constant. Since changing $\rho\to -\rho$ in
$U$ of Eq.~(\ref{eq:thetaDotD}) is equivalent to changing $\theta\to
\pi+\theta$, below we can consider only positive $\rho$. 

\begin{figure}
\includegraphics[width=1\columnwidth]{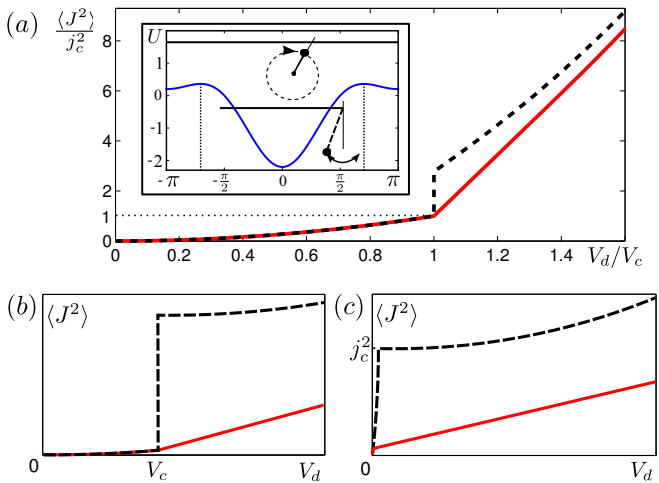} 
\caption{(color online) (a) Minimal power of Ohmic losses
  $\overline{p}=\langle J^2 \rangle /j^2_c$ as a function of drift
  velocity $V_d$ shown by solid line for $a=0.5$. The dashed line
  depicts $\overline{p}$ for dc current. The inset shows the potential
  $U(\theta)$ in which a ``particle'' is moving in the bounded
  (pendulum-like) and unbounded (rotational) regimes. A sketch of
  $\langle J^2 \rangle (V_d)$ shown by solid line in (b) for $\beta\gg
  \alpha$ ($a\ll 1$) and (c) for $\beta\ll \alpha$ ($a\gg 1$).}
\label{fig:Power_v} 
\end{figure}

Eq.~(\ref{eq:thetaDotD}) shows that there are two different regimes:
1) the bounded regime where $d< \rm{max}[U(\theta,\rho)]$ in which
case $\theta$ is bounded, and the particle oscillates in potential
well $U(\theta)$, see inset of Fig.~\ref{fig:Power_v} (a); and 2) the
rotational regime where $d> \rm{max}[U(\theta,\rho)]$ with freely
rotating magnetization in the DW.

In the bounded regime the particle moves between the two turning
points $-\theta_0$ and $\theta_0$ given by $d=U(\pm \theta_0,\rho)$.
Since $\theta$ is a bounded function $\langle\dot{\theta}\rangle=0$
and $v_{d} =-\langle\cos\theta\rangle$.  One can
show\cite{Tretiakov:losses} that in this regime the power of Ohmic
losses is minimal for dc current, i.e., $\overline{p}=v_d^2$.

In the rotational regime the term in Eq.~(\ref{eq:VdAve}) with
$\langle\dot{\theta}\rangle$ should be kept because $\theta$ is not
bounded. The equation of motion is the same as for a nonlinear
oscillator.\cite{Tretiakov:losses} Using the minimization condition
$\partial \overline{p}/\partial \rho |_{v_d}=0$ one finds
\begin{equation}
\int_{-\pi}^{\pi } \!\sqrt{d-U(\theta ,\rho )}d\theta =2\pi a\rho. 
\label{eq:min_condition}
\end{equation}
This equation defines the relationship between $d$ and $\rho$.  

The results for the minimal power of Ohmic losses $\overline{p}(v_d)$
are presented in Fig.~\ref{fig:Power_v}.  For $a>1$ there is a
critical velocity $v_{\rm{rc}}<1$, such that at $v_d<v_{\rm{rc}}$ the
power of Ohmic losses is $\overline{p}=v_d^2=p_{\rm{dc}}$. Above
$v_{\rm{rc}}$ one can minimize the Ohmic losses by moving DW with
resonant current pulses. Right above $v_{\rm{rc}}$ there is a certain
range of $v_d$ where $\overline{p}=2\rho_0 v_d -\rho_0^2$ with
$\rho_0(a)<1$ given by Eq.~(\ref{eq:min_condition}) with $d=
\rho^2$. The critical velocity is found as $v_{\rm{rc}}=\rho_0(a)$.
For $a<1$, see e.g. Fig.~\ref{fig:Power_v} (a), we find that
$v_{\rm{rc}}=1$, whereas at $v_d>1$ minimal power $\overline{p}$ is
significantly lower than $p_{\rm{dc}}$. Immediately above $v_d=1$ we
find that there is a range of $v_d$ where $\overline{p}$ is linear in
$v_d$.  At large $v_d$ the minimal power is always smaller than
$p_{\rm{dc}}$, the difference between them then approaches
$p_{\rm{dc}}-\overline{p} =(1-1/a)^2/2$.

We note that even in the limiting cases of the systems with weak
($\beta\ll \alpha$) or strong ($\beta\gg \alpha$) non-adiabatic spin
transfer torque, see Fig.~\ref{fig:Power_v} (b) and (c), where the
power of Ohmic losses is high for dc currents, the optimized ac
current gives dramatic reduction in heating power thus greatly
expanding the range of materials which can be used for spintronic
devices.  \cite{Allwood02, Parkin08} We also note that DMI suppresses
critical current $j_c$ and affects parameter $a$.

\begin{figure}
\includegraphics[width=0.9\columnwidth]{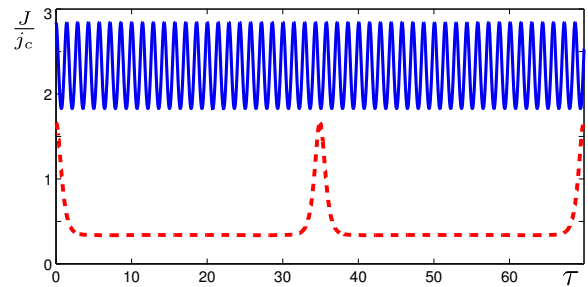} 
\caption{(color online) Resonant time-dependent current $J (\tau)$
  with $\tau=Cj_c t$ for drift velocities $v_d =0.5$ (dashed line) and
  $v_d =4.5$ (solid line) for $a=2$.}
\label{fig:currents_rotation} 
\end{figure}

For $v_d<v_{\rm{rc}}$ the optimal current coincides with the dc
current, above $v_{\rm{rc}}$ the resonant current $j(t)$ is plotted in
Fig.~\ref{fig:currents_rotation} for $a=2$ and two different
velocities $v_d$. At $v_d>v_{\rm{rc}}$ the current's maximum
$j_{\rm{max}}$ increases from $2-v_{\rm{rc}}$ at small enough
$v_d\lesssim 1$ up to $j_{\rm{max}}\approx v_d/a$ at $v_d\gg 1$.  The
current's minimum increases monotonically from small positive values
$j_{\rm{min}}= v_{\rm{rc}}$ at $v_d\sim 1$ up to
$j_{\rm{min}}=j_{\rm{max}} -2|1-a|/a$ at $v_d \gg 1$. At $v_d\lesssim
1$ (for $a>1$) the time between the current picks decreases with
increasing velocity as $T \simeq (\pi a-2\arcsin v_{\rm{rc}})/(v_d
-v_{\rm{rc}})$, whereas the pick's width is given by $\approx
1.3/\sqrt{(1-v_{\rm{rc}})}$. Therefore, at small $v_d-v_{\rm{rc}}$ the
picks are widely separated, then as $v_d$ increases the time between
the picks decreases. At $v_d \gg 1$ the optimal current has a large
constant component and small-amplitude ac modulations on top of it.

\textit{Conclusions.}  We have studied the current driven DW dynamics
in thin ferromagnetic wires. The ultimate lower bound for the Ohmic
losses in the wire has been found for any DW drift velocity $V_d$. We
have obtained the explicit time-dependence of the current which
minimizes the Ohmic losses. We believe that the use of these resonant
current pulses instead of dc current can help to dramatically reduce
heating of the wire for any $V_d$.

We thank Jairo Sinova for valuable discussions.  This work was
supported by the NSF Grant No. 0757992 and Welch Foundation (A-1678).

\bibliography{magnetizationDynamics}

\end{document}